\documentclass[envcountsame,runningheads]{llncs}

\usepackage{graphicx}
\usepackage{latexsym}
\usepackage{amsmath,amssymb,amstext}
\usepackage{comment}
\usepackage{pifont}

\usepackage{enumerate}
\usepackage{color}
\usepackage{wrapfig}
\usepackage{mathtools}

\makeatletter
\def\moverlay{\mathpalette\mov@rlay}
\def\mov@rlay#1#2{\leavevmode\vtop{%
   \baselineskip\z@skip \lineskiplimit-\maxdimen
   \ialign{\hfil$#1##$\hfil\cr#2\crcr}}}
\makeatother

\newcommand{\appsection}[1]{
  \let\oldthesection\thesection
  \renewcommand{\thesection}{Appendix \oldthesection}
  \section{#1}
  \let\thesection\oldthesection
}

\newcommand{\ignore}[1]{}

\renewcommand{\mit}{\mathit}

\newcommand{\mcl}{\mathcal}
\newcommand{\msf}{\mathsf}

\newcommand{\SN}{\msf{SN}}

\newcommand{\SNi}{\SN^\infty}
\newcommand{\SNir}[1]{\SNi_{#1}}

\newcommand{\SNirs}[1]{\funap{\SNi_{#1}}}

\newcommand{\WN}{\msf{WN}}

\newcommand{\WNi}{\WN^\infty}
\newcommand{\WNir}[1]{\WNi_{#1}}

\newcommand{\WNirs}[1]{\funap{\WNi_{#1}}}

\newcommand{\sred}{{\rightarrow}}
\newcommand{\red}{\mathrel{\sred}}

\newcommand{\sinfred}{{\threeheadrightarrow}}
\newcommand{\infred}{\mathrel{\sinfred}}

\newcommand{\pairlft}{{\langle}}                
\newcommand{\pairrgt}{{\rangle}}                
\newcommand{\pairsep}{{,\,}}                    
\newcommand{\pairstr}[1]{\pairlft#1\pairrgt}    
\newcommand{\pair}[2]{\pairstr{#1\pairsep#2}}   
\newcommand{\triple}[2]{\pair{#1\pairsep#2}}    %
\newcommand{\quadruple}[2]{\triple{#1\pairsep#2}} 

\newcommand{\sfunin}{{:}}
\newcommand{\funin}{\mathrel{\sfunin}}

\newcommand{\setemp}{{\varnothing}}

\newcommand{\mybind}[3]{#1#2.\:#3}
\newcommand{\myex}{\mybind{\exists}}
\newcommand{\myall}{\mybind{\forall}}

\newcommand{\nat}{\mathbb N}
\newcommand{\zz}{\mathbb Z}

\catcode`\@=11

\newcommand{\mywash}[2]{\setbox0=\hbox{$\m@th#1{#2}$}\wd0=0pt\box0}
\catcode`\@=12

\newcommand{\asig}{\Sigma}

\newcommand{\ster}{\mit{Ter}}

\newcommand{\siter}{\ster^{\infty}}
\newcommand{\iter}{\funap{\siter}}

\newcommand{\atrs}{R}
\newcommand{\btrs}{S}

\newcommand{\binap}[3]{#2\mathbin{#1}#3}
\newcommand{\funap}[2]{#1(#2)}

\newcommand{\bfunap}[3]{\funap{#1}{#2,#3}}
\newcommand{\tfunap}[4]{\funap{#1}{#2,#3,#4}}

\newcommand{\where}{\mathrel{|}}

\newcommand{\sdefdby}{{:}{=}}
\newcommand{\defdby}{\mathrel{\sdefdby}}

\newcommand{\threeheadrightarrow}{{\twoheadrightarrow\hspace*{-2ex}\hspace*{0.3pt}\twoheadrightarrow}}

\newcommand{\autstates}{Q}

\newcommand{\astate}{q}

\newcommand{\strff}{\msf} 

\newcommand{\lstlength}[1]{|#1|}

\newcommand{\sstrcns}{\strff{:}}

\newcommand{\strcnsd}[1]{\binap{\sstrcns}}

\newcommand{\punc}[1]{\:\text{#1}}

\newcommand{\cpi}[2]{\mathrm{\Pi}^{#1}_{#2}}
\newcommand{\csig}[2]{\mathrm{\Sigma}^{#1}_{#2}}

\newcommand{\tm}{\msf{M}}
\newcommand{\tmstates}{Q}
\newcommand{\tmsig}{\Gamma}
\newcommand{\stmtrans}{\delta}
\newcommand{\tmtrans}{\bfunap{\stmtrans}}
\newcommand{\tmiblank}{\triangleright}

\newcommand{\stmblank}{\Box}
\newcommand{\tmblank}{\funap{\stmblank}}
\newcommand{\tmL}{L}
\newcommand{\tmR}{R}
\newcommand{\stmmap}{\phi}
\newcommand{\tmmap}{\funap{\stmmap}}

\newcommand{\tmstart}{\astate_0}
\newcommand{\stmtape}{\mit{tape}}
\newcommand{\tmtape}{\funap{\stmtape}}
\newcommand{\tmconf}[1]{\mathcal{C}\!\mit{onf}_{#1}}

\newcommand{\tmfinal}[2]{\funap{\mit{final}_{#2}}{{#1}}}

\newcommand{\stmzer}{\msf{0}}
\newcommand{\tmzer}{\funap{\stmzer}}
\newcommand{\stmsucc}{\msf{S}}

\newcommand{\tmT}{\msf{T}}

\newcommand{\stmpeb}{\bullet}
\newcommand{\tmpeb}{\funap{\stmpeb}}

\newcommand{\tmtrs}[1]{\atrs_{#1}}
\newcommand{\tmtrspeb}[1]{\atrs_{#1}^{\stmpeb}}
\newcommand{\tmstep}{\to_\tm}

\newcommand{\stmfun}[1]{f_{#1}}
\newcommand{\tmfun}[1]{\funap{\stmfun{#1}}}
\newcommand{\tmrel}[1]{\mathrel{>_{#1}}}

\newcommand{\sfs}{\stmsucc}
\newcommand{\fs}{\funap{\sfs}}
\newcommand{\sfc}{\msf{c}}
\newcommand{\fc}{\funap{\sfc}}

\newcommand{\pickn}{\msf{pickn}}
\newcommand{\picknok}{\funap{\msf{ok}}}
\newcommand{\trspickn}{\atrs_{\pickn}}

\newcommand{\pto}{\rightharpoonup}

\newcommand{\weg}[1]{}
\def\phi{\varphi}

\begin{document}

\title{Levels of Undecidability in Infinitary Rewriting:\\ Normalization and Reachability}

\author{
  J\"{o}rg Endrullis
}

\institute{
    Free University Amsterdam, The Netherlands\\
    \email{joerg@few.vu.nl}
}
\maketitle

\begin{abstract}
  In \cite{EGZ09} it has been shown that infinitary strong normalization ($\SNi$) is $\cpi{1}{1}$-complete. 
  Suprisingly, it turns out that infinitary weak normalization ($\WNi$) is a harder problem, 
  being $\cpi{1}{2}$-complete, and thereby strictly higher in the analytical hierarchy.
\end{abstract}

We assume familiarity with infinitary term rewriting;
we further reading we refer to~\cite{terese:03,klop:vrij:05}.

\section{Infinitary Strong Normalization and Reachability}\label{sec:isn}

\begin{definition}\normalfont\label{def:tm}
  A \emph{Turing machine} $\tm$ is a quadruple $\quadruple{\tmstates}{\tmsig}{\tmstart}{\stmtrans}$
  consisting of:
  \begin{itemize}
  \item finite set of states $\tmstates$,
  \item an initial state $q_0 \in \tmstates$,
  \item a finite alphabet $\tmsig$ containing a designated symbol $\stmblank$, called \emph{blank}, and
  \item a partial \emph{transition function} 
        $\stmtrans \funin \autstates \times \tmsig \to \tmstates \times \tmsig \times \{\tmL,\tmR\}$.
  \end{itemize}
  A \emph{configuration} of a Turing machine is a pair $\pair{\astate}{\stmtape}$
  consisting of a state $\astate \in \tmstates$
  and the tape content $\stmtape \funin \zz \to \tmsig$
  such that the carrier $\{n \in \zz \where \tmtape{n} \ne \stmblank\}$ is finite.
  The set of all configurations is denoted $\tmconf{\tm}$.
  We define the relation $\tmstep$ on the set of configurations $\tmconf{\tm}$ as follows:
  $\pair{\astate}{\stmtape} \tmstep \pair{\astate'}{\stmtape'}$
  whenever:
  \begin{itemize}
   \item
     $\tmtrans{\astate}{\tmtape{0}} = \triple{\astate'}{f}{\tmL}$,
     $\funap{\stmtape'}{1} = f$ and $\myall{n \ne 0}{\funap{\stmtape'}{n+1} = \tmtape{n}}$, or
   \item
     $\tmtrans{\astate}{\tmtape{0}} = \triple{\astate'}{f}{\tmR}$,
     $\funap{\stmtape'}{-1} = f$ and $\myall{n \ne 0}{\funap{\stmtape'}{n-1} = \tmtape{n}}$.
  \end{itemize}
\end{definition}

Without loss of generality we assume that $\tmstates \cap \tmsig = \setemp$,
that is, the set of states and the alphabet are disjoint.
This enables us to denote configurations
as $\triple{w_1}{\astate}{w_2}$,
denoted $w_1^{-1} \astate w_2$ for short,
with $w_1,w_2 \in \tmsig^*$ and $\astate \in \tmstates$,
which is shorthand for $\pair{\astate}{\stmtape}$
where $\tmtape{n} = \funap{w_2}{n+1}$ for $0 \le n < \lstlength{w_2}$,
and $\tmtape{-n} = \funap{w_1}{n}$ for $1 \le n \le \lstlength{w_1}$
and $\tmtape{n} = \stmblank$ for all other positions $n \in \zz$.

The Turing machines we consider are deterministic.
As a consequence, final states are unique (if they exist),
which justifies the following definition.

\begin{definition}\normalfont
  Let $\tm$ be a Turing machine and $\pair{\astate}{\stmtape} \in \tmconf{\tm}$.
  We denote by $\tmfinal{\pair{\astate}{\stmtape}}{\tm}$
  the $\tmstep$-normal form of $\pair{\astate}{\stmtape}$
  if it exists and undefined, otherwise.
  Whenever $\tmfinal{\pair{\astate}{\stmtape}}{\tm}$ exists
  then we say that \emph{$\tm$ halts on $\pair{\astate}{\stmtape}$
  with final configuration $\tmfinal{\pair{\astate}{\stmtape}}{\tm}$}.
  Furthermore we say \emph{$\tm$ halts on $\stmtape$} as shorthand for
  \emph{$\tm$ halts on $\pair{\tmstart}{\stmtape}$}.
\end{definition}

Turing machines can compute $n$-ary functions $f \funin \nat^n \to \nat$
or relations $S \subseteq \nat^*$. 
We need only unary functions $\stmfun{\tm}$ and binary ${\tmrel{\tm}} \subseteq \nat \times \nat$ relations.

\begin{definition}\label{def:funcrel}\normalfont
  Let $\tm = \quadruple{\tmstates}{\tmsig}{\tmstart}{\stmtrans}$ be a Turing machine with $\stmsucc,\stmzer \in \tmsig$.
  We define a partial function $\stmfun{\tm} \funin \nat \pto \nat$ for all $n \in \nat$ by:
  \begin{align*}
  \tmfun{\tm}{n}
  = \begin{cases}
      m & \text{if }
      \tmfinal{\tmstart \stmsucc^{n} \stmzer}{\tm}
      = \ldots \astate \stmsucc^m \stmzer \ldots\\
      \text{undefined} & \text{otherwise}
    \end{cases}
  \end{align*}
  and for $\tm$ total (i.e.\ $\tm$ halts on all tapes) we define the binary relation ${\tmrel{\tm}} \subseteq \nat \times \nat$ by:
  \begin{align*}
  n \tmrel{\tm} m \;\Longleftrightarrow\;
    \tmfinal{\stmzer \stmsucc^{n} \tmstart \stmsucc^{m} \stmzer}{\tm}
    = \ldots \astate \stmzer \ldots
  \punc.
  \end{align*}
\end{definition}
Note that, the set $\{\,\tmrel{\tm} \where \tm \text{ a Turing machine that halts on all tapes}\,\}$
is the set of recursive binary relations on $\nat$.

We use the translation of Turing machines $\tm$ to TRSs $\tmtrs{\tm}$ from \cite{jw:handbook}.
\begin{definition}\normalfont\label{def:tmtrs}
  For every Turing machine $\tm = \quadruple{\tmstates}{\tmsig}{\tmstart}{\stmtrans}$
  we define a TRS $\tmtrs{\tm}$ as follows.
  The signature is $\asig = \tmstates \cup \tmsig \cup \{\tmiblank\}$
  where the symbols $\astate \in \tmstates$ have arity 2,
  the symbols $f \in \tmsig$ have arity 1
  and $\tmiblank$ is a constant symbol, 
  which represents an infinite number of blank symbols.
  The rewrite rules of $\tmtrs{\tm}$ are:
  \begin{align*}
    \bfunap{\astate}{x}{\funap{f}{y}} &\to \bfunap{\astate'}{\funap{f'}{x}}{y}
    &&\text{ for every }\tmtrans{\astate}{f} = \triple{\astate'}{f'}{\tmR}\\
    \bfunap{\astate}{\funap{g}{x}}{\funap{f}{y}} &\to \bfunap{\astate'}{x}{\funap{g}{\funap{f'}{y}}}
    &&\text{ for every }\tmtrans{\astate}{f} = \triple{\astate'}{f'}{\tmL}
  \end{align*}
  together with four rules for `extending the tape':
  \begin{align*}
    \bfunap{\astate}{\tmiblank}{\funap{f}{y}} &\to \bfunap{\astate'}{\tmiblank}{\tmblank{\funap{f'}{y}}}
    &&\text{ for every }\tmtrans{\astate}{f} = \triple{\astate'}{f'}{\tmL}\\
    \bfunap{\astate}{x}{\tmiblank} &\to \bfunap{\astate'}{\funap{f'}{x}}{\tmiblank}
    &&\text{ for every }\tmtrans{\astate}{\stmblank} = \triple{\astate'}{f'}{\tmR}\\
    \bfunap{\astate}{\funap{g}{x}}{\tmiblank} &\to \bfunap{\astate'}{x}{\funap{g}{\funap{f'}{\tmiblank}}}
    &&\text{ for every }\tmtrans{\astate}{\stmblank} = \triple{\astate'}{f'}{\tmL}\\
    \bfunap{\astate}{\tmiblank}{\tmiblank} &\to \bfunap{\astate'}{\tmiblank}{\tmblank{\funap{f'}{\tmiblank}}}
    &&\text{ for every }\tmtrans{\astate}{\stmblank} = \triple{\astate'}{f'}{\tmL}
    \punc.
  \end{align*}
\end{definition}

In~\cite{EGZ09} the TRSs $\tmtrs{\tm}$ has been extended as follow to prove $\cpi{1}{1}$-completeness
of finiteness of dependency pair problems:
\begin{definition}[\cite{EGZ09}]\normalfont\label{def:tmtrs-adap}
  For every Turing machine $\tm = \quadruple{\tmstates}{\tmsig}{\tmstart}{\stmtrans}$
  we define the TRS $\tmtrspeb{\tm}$ as follows.
  The signature $\asig = \tmstates \cup \tmsig \cup \{\tmiblank,\stmpeb,\tmT\}$ where $\stmpeb$ is a unary symbol,
  $\tmT$ is a constant symbol,
  and the rewrite rules of $\tmtrspeb{\tm}$ are:
  \begin{align*}
    \ell &\to \tmpeb{r}
    &&\text{ for every }\ell \to r \in \tmtrs{\tm}
  \end{align*}
  and rules for rewriting to $\tmT$ after successful termination:
  \begin{align*}
    \bfunap{\astate}{x}{\tmzer{y}} &\to \tmT
    &&\text{ whenever }\tmtrans{\astate}{\stmsucc}\text{ is undefined}\\
    \tmpeb{\tmT} &\to \tmT
    \punc.
  \end{align*}
Moreover, we define the TRS $\trspickn$ to consist of the following rules:
  \begin{align*}
    \pickn &\to \fc{\pickn} &
    \pickn &\to \picknok{\tmzer{\tmiblank}} &
    \fc{\picknok{x}} &\to \picknok{\fs{x}}
    \punc.
  \end{align*}
\end{definition}

\begin{proposition}\label{prop:snitrs}
  Let $\tm$ be an arbitrary Turing machine. 
  We define the TRS $\btrs$ together to consist of the rules of $\tmtrspeb{\tm} \uplus \trspickn$
  together with:
  \begin{align}
    \tfunap{\msf{run}}{\tmT}{\picknok{x}}{\picknok{y}} &\to \tfunap{\msf{run}}{\bfunap{\tmstart}{x}{y}}{\picknok{y}}{\pickn}
    \punc,\label{rule:dp}
  \end{align}
  and define a term $t \defdby \tfunap{\msf{run}}{\tmT}{\pickn}{\pickn}$.
  Then it holds:
  \[\SNir{\btrs} \Longleftrightarrow \SNirs{\btrs}{t} \Longleftrightarrow {\tmrel{\tm}} \text{ is well-founded}\punc.\]
\end{proposition}

\begin{proof}
  See~\cite{EGZ09}.\qed
\end{proof}

\begin{theorem}\label{thm:sni}
  Uniform infinitary strong normalization, $\SNir{\atrs}$, and for single terms, $\SNirs{\atrs}{s}$ is $\cpi{1}{1}$-complete.
\end{theorem}
\begin{proof}
  The $\cpi{1}{1}$-hardness has been shown in~\cite{EGZ09} using
  that well-foundedness is $\cpi{1}{1}$-complete.

  It remains to be shown that the property is in $\cpi{1}{1}$
  (in~\cite{EGZ09} this has been done only reductions of length $\omega$).
    A finite or infinite term $t$ can be encoded as a function $t \funin \nat \to \nat$
  (from positions to symbols from the signature).
  An infinite reduction can be rendered as a function $\sigma \funin \alpha \to ((\nat \to \nat) \times \nat)$
  from an ordinal $\alpha$ to terms together with the rewrite position (here we assume that an ordinal is the set of all smaller ordinals)
  where $\funap{\sigma}{\beta}$ is the $\beta$-th term of the sequence together with the rewrite position,
  and we require:
  \begin{enumerate}
    \item $\funap{\sigma}{\beta}$ rewrites to $\funap{\sigma}{\beta+1}$ for all $\beta < \alpha$, and
    \item for all limit ordinals $\beta < \alpha$, and $\gamma$ approaching $\beta$ from below, we have:
    \begin{itemize}
      \item $\funap{\sigma}{\gamma}$ converges to $\funap{\sigma}{\beta}$, and
      \item the depth of the $\gamma$-th rewrite steps tends to infinity.
    \end{itemize}
  \end{enumerate}
  If condition~(ii) holds for all limit ordinals $\beta \le \alpha$ then the rewrite sequence $\sigma$ is called strongly convergent.
  An ordinal $\alpha$ can be viewed as a well-founded relation $\alpha \subseteq \nat \times \nat$.
  The property of a relation to be well-founded can be expressed by a $\cpi{1}{1}$-formula,
  and the above properties on rewrite sequences are arithmetic.
  By~\cite{klop:vrij:05} the property $\SNirs{\atrs}{s}$ holds if and only if
  all reductions admitted by $s$ are strongly convergent.
  Hence $\SNir{\atrs}$ and $\SNirs{\atrs}{s}$ can be expressed by a $\cpi{1}{1}$-formula 
  since the above conditions~(i) and~(ii) are arithmetic.
  \qed
\end{proof}

Using a minor modification of the term rewriting system from Proposition~\ref{prop:snitrs}
we obtain that weak normalization for single terms and \emph{reachability} are $\csig{1}{1}$-complete, that is,
the problem of deciding on the input of a TRS $S$ and terms $s$, $t$ whether $s \infred_S t$.

\begin{theorem}
  Infinitary weak normalization for single terms, $\WNirs{\atrs}{s}$,
  and reachability in infinitary rewriting
  are $\csig{1}{1}$-complete.
\end{theorem}
\begin{proof}
  Let $\tm$ be an arbitrary Turing machine. 
  We define the TRS $\btrs'$ together to consist of the rules of $\tmtrspeb{\tm} \uplus \trspickn$
  together with:
  \begin{align}
    \tfunap{\msf{run}}{\tmT}{\picknok{x}}{\picknok{y}} &\to \tmpeb{\tfunap{\msf{run}}{\bfunap{\tmstart}{x}{y}}{\picknok{y}}{\pickn}}
    \punc,
  \end{align}
  and define a term $t \defdby \tfunap{\msf{run}}{\tmT}{\pickn}{\pickn}$.
  We have $t \infred_{\btrs'} \stmpeb^\infty$ if and only if $t$ admits
  a rewrite sequence containing infinitely many root steps with respect to
  the rewrite system $\btrs$ from Proposition~\ref{prop:snitrs}.
  As a consequence we have:
  \[t \infred_{\btrs'} \stmpeb^\infty \Longleftrightarrow \neg \SNirs{\btrs}{t} \Longleftrightarrow {\tmrel{\tm}} \text{ is not well-founded}\punc. \]
  Hence reachability is $\csig{1}{1}$-hard.
  
  We add one more rule to $\btrs'$:
  \begin{align}
    \tfunap{\msf{run}}{x}{y}{z} &\to \tfunap{\msf{run}}{x}{y}{z}
    \punc,
  \end{align}
  Note that this rule has no impact on reachability.
  Then $\WNirs{\btrs'}{t}$ holds if and only if $t \infred_{\btrs'} \stmpeb^\infty$,
  and hence $\WNirs{\btrs'}{t}$ is $\csig{1}{1}$-complete.
  
  Moreover, weak normalization for single terms and reachability are in $\csig{1}{1}$.
  We have $\WNirs{\btrs'}{t}$ if and only if there exists a normal form $t'$
  such that $t \infred t'$,
  and we have reachability $s \infred t$ if and only if there exists
  a reduction from $s$ to $t$.
  The quantification over terms and rewrite sequences are existential set or function quantifiers
  (which can be compressed to one single quantifier), and
  all other properties are arithmetic;
  see the encoding of reduction sequences see the proof of Theorem~\ref{thm:sni}.
  \qed
\end{proof}

\section{Uniform Infinitary Weak Normalization}\label{sec:iwn}
\newcommand{\tmsrs}[1]{S_{#1}}
\newcommand{\pow}[1]{\funap{\mathcal{P}}{#1}}
\renewcommand{\stmtape}{\sigma}
\renewcommand{\tmtape}{\funap{\stmtape}}
\newcommand{\prefix}[2]{#1_{<#2}}
\newcommand{\suffix}[2]{#1_{\ge#2}}
\newcommand{\sequence}[2]{{\{#1\}}_{#2}}
\newcommand{\arun}{r}
\newcommand{\olang}{\funap{\mcl{L}^\omega}}

For $\stmtape \in \tmsig^\infty$ and $i \in \nat$ 
we write $\prefix{\stmtape}{i}$ for the prefix of $\stmtape$ up to (excluding) position $i$,
and
$\suffix{\stmtape}{i}$ for the suffix of $\stmtape$ starting from (including) position $i$.
We define non-deterministic Turing machines with one-sided infinite tape.

\begin{definition}\normalfont
  A \emph{non-deterministic (one-sided) Turing machine} $\tm$ 
  is a quadruple $\quadruple{\tmstates}{\tmsig}{\tmstart}{\stmtrans}$
  consisting of:
  \begin{itemize}
  \item finite set of states $\tmstates$,
  \item an initial state $q_0 \in \tmstates$,
  \item a finite alphabet $\tmsig$ containing a designated symbol $\stmblank$, called \emph{blank}, and
  \item a partial \emph{transition function} 
        $\stmtrans \funin \autstates \times \tmsig \to \pow{\tmstates \times \tmsig \times \{\tmL,\tmR\}}$.
  \end{itemize}
  A \emph{configuration} of $\tm$ is a triple $\triple{\astate}{\stmtape}{i}$
  consisting of a state $\astate \in \tmstates$,
  a tape content $\stmtape \funin \tmsig^\omega$,
  and the position of the head $i \in \nat$.
 
  For two configurations we define 
  $\triple{\astate}{\stmtape}{i} \tmstep \triple{\astate'}{\stmtape'}{i'}$
  whenever:
  \begin{itemize}
   \item
     $\triple{\astate'}{f}{\tmL} \in \tmtrans{\astate}{\tmtape{i}}$, $i > 0$
     $i' = i-1$, and $\stmtape' = \prefix{\stmtape}{i} \; f \; \suffix{\stmtape}{i+1}$
   \item
     $\triple{\astate'}{f}{\tmR} \in \tmtrans{\astate}{\tmtape{i}}$,
     $i' = i+1$, and $\stmtape' = \prefix{\stmtape}{i} \; f \; \suffix{\stmtape}{i+1}$
  \end{itemize}
  An infinite sequence of configurations $\arun \funin \sequence{\triple{\astate_j}{\stmtape_j}{i_j}}{j \ge 0}$
  is a \emph{run of $\tm$ on $\stmtape$} if:
  \begin{enumerate}
    \item $\triple{\astate_0}{\stmtape_0}{i_0} = \triple{q_0}{\stmtape}{0}$, and
    \item $\triple{\astate_j}{\stmtape_j}{i_j} \tmstep \triple{\astate_{j+1}}{\stmtape_{j+1}}{i_{j+1}}$ for all $j \ge 0$.
  \end{enumerate}
  A run $\arun \funin \sequence{\triple{\astate_j}{\stmtape_j}{i_j}}{j \ge 0}$ is called \emph{complete}
  if every position is visited, that is, $\myall{n \ge 0}{\myex{j \ge 0}{i_j = n}}$,
  and $\arun$ is called \emph{oscillating} if $\myex{n \ge 0}{\myall{j \ge 0}{\myex{j' > j}{i_{h'} = n}}}$.
\end{definition}

\begin{definition}\normalfont
  A run is called \emph{accepting} if it is complete and non-oscillating.
  The \emph{$\omega$-language} $\olang{\tm}$ accepted by a non-deterministic Turing machine $\tm$ is:
  \begin{align*}
    \olang{\tm} = \{w \in \tmsig^\omega \mid \emph{there exists an accepting run of $\tm$ on $w$} \}
  \end{align*}
\end{definition}

Notice that accepting runs visits every symbol at least once, but only finitely often.
The following is a proposition from \cite{omega:computations}:
\begin{proposition}[\cite{omega:computations}]
  The set $\{\tm \mid \olang{\tm} = \tmsig^\omega\}$ is $\cpi{1}{2}$-complete. \qed
\end{proposition}

We use the translation of Turing machines $\tm$ to string rewriting systems $\tmsrs{\tm}$ from \cite{terese:03}.
\begin{definition}\normalfont
  For every (non-deterministic) Turing machine $\tm = \quadruple{\tmstates}{\tmsig}{\tmstart}{\stmtrans}$
  we define a TRS $\tmsrs{\tm}$ as follows.
  The signature $\asig$ consists of symbols from $\tmstates \cup \tmsig$
  all having arity 1.
  The rewrite rules of $\tmsrs{\tm}$ are:
  \begin{align*}
    \funap{\astate}{\funap{f}{x}} &\to \funap{f'}{\funap{\astate'}{x}}
    &&\text{ for every }\triple{\astate'}{f'}{\tmR} \in \tmtrans{\astate}{f}\\
    \funap{g}{\bfunap{\astate}{\funap{f}{x}}} &\to \funap{\astate'}{\funap{g}{\funap{f'}{x}}}
    &&\text{ for every }\triple{\astate'}{f'}{\tmL} \in \tmtrans{\astate}{f}
  \end{align*}
\end{definition}

\begin{definition}\label{def:tmnmap}
  Let $\tm$ be a non-deterministic Turing machine.
  We define a mapping $\stmmap \funin (\tmsig \cup \tmstates)^* \to \iter{\tmsig \cup \tmstates}$
  by $\tmmap{a\, w} \defdby \funap{a}{\tmmap{t}}$,
  and we extend $\stmmap$ to configurations $\triple{\astate}{\stmtape}{i}$ of $\tm$
  by defining: $\tmmap{\triple{\astate}{\stmtape}{i}} = \tmmap{\prefix{\stmtape}{i} \,\astate\, \suffix{\stmtape}{i}}$.
\end{definition}

The following lemma follows immediately from the definition of $\tmsrs{\tm}$:
\begin{lemma}\label{lem:tmnmap}
  Let $\tm$ be a non-deterministic Turing machine.
  For configurations $c_1$, $c_2$ of $\tm$
  we have $c_1 \tmstep c_2$ if and only if $\tmmap{c_1} \red_{\tmsrs{\tm}} \tmmap{c_2}$.\qed
\end{lemma}

Then we obtain the following lemma establishing a correspondence of
strongly convergent rewrite sequences and complete, non-oscillating runs:
\begin{lemma}\label{lem:ndet}
  Let $\tm$ be a non-deterministic Turing machine and $\stmtape \in \tmsig^\omega$. 
  Then $\stmtape \in \olang{\tm}$ if and only if
  $\funap{q_0}{\tmmap{\stmtape}} \infred_{\tmsrs{\tm}} t$ for some ground term $t \in \iter{\tmsig}$,
  that is, $t = \tmmap{\stmtape'}$ for some $\stmtape' \in \tmsig^\omega$.
\end{lemma}
\begin{proof}
  By Lemma~\ref{lem:tmnmap} every rewrite sequence $\tmmap{\triple{q_0}{\stmtape}{0}} \infred \ldots$ corresponds
  to a run of $\tm$ on $\stmtape$.
  By definition of $\infred$ the limit term exists 
  if and only if the rewrite sequence is strongly convergent
  and this holds 
  if and only if every rewrite position occurs at only finitely often,
  that is, the run is complete and non-oscillating.\qed
\end{proof}

\begin{theorem}
  Uniform infinitary weak normalization, $\WNir{\atrs}$, is $\cpi{1}{2}$-complete.
\end{theorem}
\newcommand{\run}[4]{\funap{\msf{run}}{#1,#2,#3,#4}}
\newcommand{\scomp}[1]{\Delta_{#1}}
\newcommand{\comp}[2]{\funap{\scomp{#1}}{#2}}
\begin{proof}
  Let $\tm$ be a non-deterministic Turing machine.
  We define the TRS $R$ as an extension of the TRS $\tmsrs{\tm}$ with the following rules:
  \begin{align}
    \run{x}{y}{z}{z} &\to \run{\xi}{\funap{q_0}{z}}{\comp{1}{z}}{\comp{1}{z}} \label{rule:ndet1}\\
    \run{x}{x}{y}{z} &\to \bot\label{rule:stop}\\
    \funap{\astate}{x} &\to \bot &&\text{for $\astate \in \tmstates$}\\
    \xi &\to \funap{f}{\xi} &&\text{for all $f \in \tmsig$}\\
    \comp{1}{\funap{f}{x}} &\to \funap{f}{\comp{1}{x}} &&\text{for all $f \in \tmsig$}\\
    \comp{2}{\funap{f}{x}} &\to \funap{f}{\comp{2}{x}} &&\text{for all $f \in \tmsig$}
  \end{align}
  The rules for $\xi$, $\scomp{1}$ and $\scomp{2}$ are obviously infinitary normalizing,
  in particular the normal forms of $\xi$ are exactly all ground terms from $\iter{\tmsig}$.
  By application of $\funap{\astate}{x} \to \bot$ every term can be rewritten to a normal form with respect to $\tmsrs{\tm}$.
  Moreover, $\comp{1}{t}$ and $\comp{2}{t}$ have a common reduct if and only if $t \infred t'$
  for a ground term from $\iter{\tmsig}$.
  
  Assume that $\olang{\tm} = \tmsig^\omega$.
  If there exists a term that is not infinitary weakly normalizing,
  then by the above considerations it must admit a rewrite sequence where 
  (at some fixed position) the first rule is applied infinitely often.
  By the shape of~\eqref{rule:ndet1} $\run{s}{t}{u}{u} \to \run{\xi}{\funap{q_0}{u}}{\comp{1}{u}}{\comp{1}{u}}$
  this can only occur if $\comp{1}{u}$ and $\comp{2}{u}$ have a common reduct.
  This implies that $u \infred u'$ for some ground term $u' \in \iter{\tmsig}$.
  Then $\funap{q_0}{u} \infred \funap{q_0}{u'}$
  and by Lemma~\ref{lem:ndet} we have 
  $\funap{q_0}{u'} \infred u''$ for some ground term $u'' \in \iter{\tmsig}$
  (note that $\funap{\stmmap^{-1}}{u'} \in \olang{\tm}$).
  Since also $\xi \infred u''$ we get
  a rewrite step of the from $\run{u''}{u''}{\xi}{u} \to \bot$.
  Hence every term is $\WNir{\tmsrs{\tm}}$.

  Assume that $\olang{\tm} \ne \tmsig^\omega$.
  Then there exists $w \in \tmsig^\omega$ for which there exists no accepting run of $\tm$.
  Let $u = \tmmap{w}$.
  We claim that the term $$\run{\xi}{\funap{q_0}{u}}{\comp{1}{u}}{\comp{2}{u}}$$
  is not infinitary weakly normalizing.
  Note that
  $\funap{q_0}{u}$ does not reduce to a term containing $\xi$, and
  the only $\xi$-free reducts of $\xi$ are terms from $\iter{\tmsig}$.
  However, $\funap{q_0}{u}$ does not reduce to a ground terms from $\iter{\tmsig}$ 
  by Lemma~\ref{lem:ndet} since there $\tm$ admits no accepting run for $w$.
  Consequently, the rule~\ref{rule:stop} is is never applicable,
  and we cannot get rid of the redexes in $\comp{1}{u}$ and $\comp{2}{u}$
  unless we reduce both to their unique normal form: 
  $\comp{1}{u} \infred u$ and $\comp{2}{u} \infred u$.
  However, then we have a root redex giving rise to a step:
  $$\run{s}{t}{u}{u} \to \run{\xi}{\funap{q_0}{u}}{\comp{1}{u}}{\comp{1}{u}}$$
  This concludes $\cpi{0}{2}$-hardness of uniform $\WNi$\!.

  It remains be shown that $\WNi$ is in $\cpi{1}{2}$.
  The property $\WNir{\atrs}$ holds if and only if 
  for all terms $s \in \iter{\asig}$ there exists a reduction to a normal form.
  (For the encoding of reduction sequences see the proof of Theorem~\ref{thm:sni}.)
  Hence $\WNi$ is in $\cpi{1}{2}$.
 \qed
\end{proof}

\bibliography{main}

\begin{thebibliography}{1}

\bibitem{omega:computations}
J.~Castro and F.~Cucker.
\newblock Nondeterministic $\omega$-computations and the analytical hierarchy.
\newblock {\em Journal Math. Logik und Grundlagen d. Math}, 35:333–--342,
  1989.

\bibitem{EGZ09}
J.~Endrullis, H.~Geuvers, and H.~Zantema.
\newblock Degrees of undecidability in term rewriting.
\newblock In E.~Gr{\"a}del and R.~Kahle, editors, {\em Proceedings of Computer
  Science Logic (CSL09)}, volume 5771 of {\em Lecture Notes in Computer
  Science}, pages 255--270. Springer, 2009.

\bibitem{jw:handbook}
J.~W. Klop.
\newblock {Term rewriting systems}.
\newblock In S.~Abramsky, D.~M. Gabbay, and S.~E. Maibaum, editors, {\em
  Handbook of Logic in Computer Science}, volume~2, pages 1--116. Oxford
  University Press, Inc., 1992.

\bibitem{klop:vrij:05}
J.~W. Klop and R.~C.~d. Vrijer.
\newblock {Infinitary Normalization}.
\newblock In S.~Artemov, H.~Barringer, A.~d'Avila Garcez, L.~Lamb, and
  J.~Woods, editors, {\em We Will Show Them: Essays in Honour of Dov Gabbay},
  volume~2, pages 169--192. {College Publ.}, 2005.

\bibitem{terese:03}
Terese.
\newblock {\em {Term Rewriting Systems}}, volume~55 of {\em Cambridge Tracts in
  Theoretical Computer Science}.
\newblock Cambridge University Press, 2003.

\end{thebibliography}
\bibliographystyle{abbrv}

\end{document}